\documentstyle[12pt]{article}
\input html.sty
\begin{document}
\title{MATTERS OF GRAVITY, The newsletter of the APS TIG on Gravitation}
\begin{center}
{ \Large {\bf MATTERS OF GRAVITY}}\\
\medskip
\hrule
\medskip
{The newsletter of the Topical Group in Gravitation of the American Physical 
Society}\\
\medskip
{\bf Number 8 \hfill Fall 1996}
\end{center}
\begin{flushleft}
\bigskip
\bigskip
\tableofcontents
\vfill
\section*{\noindent  Editor\hfill}

\medskip
Jorge Pullin\\
\smallskip
Center for Gravitational Physics and Geometry\\
The Pennsylvania State University\\
University Park, PA 16802-6300\\
smallskip
Fax: (814)863-9608\\
Phone (814)863-9597\\
Internet: 
\htmladdnormallink{\protect {\tt pullin@phys.psu.edu}}
{mailto:pullin@phys.psu.edu}\\
WWW: \htmladdnormallink{\protect {\tt html://www.phys.psu.edu/PULLIN}}
{http://www.phys.psu.edu/PULLIN}
\begin{rawhtml}
<P>
<BR><HR><P>
\end{rawhtml}
\end{flushleft}
\pagebreak
\section*{Editorial}

Not much to say in this editorial except that the newsletter is now
produced in LaTeX, this will allow us to generate automatically an
html version of it and better distribute it through the World Wide Web
to the TeX/Postscript impaired. By the way, the html version is really
something: every occurrence of an email address, preprint archive
reference or web page is hotlinked, so you can click and surf. It also
led to one of the worst Sundays in my life: getting LaTeX to do what
you want is difficult enough without having to plan ahead your LaTeX
code to be readable by an ``intelligent translator program''
(LaTeX2html by Nikos Drakos, a wonderful tool).  

As usual I wish to again remind people that suggestions for
authors/topics for the newsletter are very welcome.

We wish to say good bye and thanks a lot to Peter Michelson and
welcome Warren Johnson as correspondent for bar-type gravitational
wave detectors.

The next newsletter is due February 1st.  If everything goes well this
newsletter should be available in the gr-qc Los Alamos archives under
number gr-qc/9609008. To retrieve it send email to 
\htmladdnormallink{gr-qc@xxx.lanl.gov}{mailto:gr-qc@xxx.lanl.gov}
(or 
\htmladdnormallink{gr-qc@babbage.sissa.it}{mailto:gr-qc@babbage.sissa.it} 
in Europe) with Subject: get 9609008
(numbers 2-7 are also available in gr-qc). All issues are available in the
WWW:\\\htmladdnormallink{\protect {\tt
http://vishnu.nirvana.phys.psu.edu/mog.html}}
{http://vishnu.nirvana.phys.psu.edu/mog.html}\\ 
A hardcopy of the newsletter is
distributed free of charge to the members of the APS
Topical Group on Gravitation. It is considered a lack of etiquette to
ask me to mail you hard copies of the newsletter unless you have
exhausted all your resources to get your copy otherwise.

If you have comments/questions/complaints about the newsletter email
me. Have fun.
\bigbreak

\hfill Jorge Pullin\vspace{-0.8cm}
\section*{Correspondents}
\begin{itemize}
\item John Friedman and Kip Thorne: Relativistic Astrophysics,
\item Raymond Laflamme: Quantum Cosmology and Related Topics
\item Gary Horowitz: Interface with Mathematical High Energy Physics and
String Theory
\item Richard Isaacson: News from NSF
\item Richard Matzner: Numerical Relativity
\item Abhay Ashtekar and Ted Newman: Mathematical Relativity
\item Bernie Schutz: News From Europe
\item Lee Smolin: Quantum Gravity
\item Cliff Will: Confrontation of Theory with Experiment
\item Peter Bender: Space Experiments
\item Riley Newman: Laboratory Experiments
\item Warren Johnson: Resonant Mass Gravitational Wave Detectors
\item Stan Whitcomb: LIGO Project
\end{itemize}
\vfill
\pagebreak
\section*{\centerline {April 1997 Joint APS/AAPT Meeting}}
\addtocontents{toc}{\protect\medskip}
\addtocontents{toc}{\bf APS Topical Group in Gravitation News:}
\addtocontents{toc}{\protect\medskip}
\addcontentsline{toc}{subsubsection}{\it  April 1997 Joint APS/AAPT Meeting}

\begin{center}

\bigskip
{CALL FOR PAPERS} (0th announcement):	
\end{center}

The 1997 Joint American Physical Society/American Association of
Physics Teachers Meeting will be held April 18-21 1997 in Washington,
DC.  This meeting will feature invited sessions sponsored by the
Topical Group in Gravitation (GTG) as well as the GTG annual business
meeting. The Ligo Research Community will also hold its meeting
here. This year the GTG will organize approximately two focus sessions
with invited and contributed talks on specific topics of interest to
the GTG membership. The details of the focus sessions will be made
available as soon as possible on the APS Meetings and GTG Web pages:\\
\htmladdnormallink{\protect {\tt
http://www.aps.org/meet/meetcal.html}}
{http://www.aps.org/meet/meetcal.html}\\ \htmladdnormallink{\protect
{\tt http://vishnu.nirvana.phys.psu.edu/tig/}}
{http://vishnu.nirvana.phys.psu.edu/tig/}

Contributed
papers are also welcomed from MOG readers on (1) experiments and
observations related to the detection and interpretation of
gravitational waves, (2) experimental tests of gravitational theories,
(3) computational general relativity, (4) relativistic astrophysics,
(5) theories of the gravitational field, solutions to the field
equations, and properties of solutions, (6) classical and quantum
cosmology, and (7) quantum gravity.

The abstract deadline (see 
\htmladdnormallink{\protect {\tt http://www.aps.org/meet/meetcal.html}}
{http://www.aps.org/meet/meetcal.html})
is
not yet posted but will probably be around the end of the year. To
submit an abstract, APS but not GTG membership is required (an APS
member may submit an abstract for a non-member). 

See the section on
electronic submission of abstracts at\\
\htmladdnormallink{\protect {\tt http://www.aps.org/meet/index.html}}
{http://www.aps.org/meet/index.html}.
\vfill
\pagebreak
\section*{\centerline {GEO600}}
\addtocontents{toc}{\protect\bigskip}
\addtocontents{toc}{\bf Research briefs:}
\addtocontents{toc}{\protect\medskip}
\addcontentsline{toc}{subsection}{\it GEO600 by Karsten Danzmann}
\begin{center}
   {\bf       Buildings and trenches finished; 
       Installation of vacuum tube beginning}\\
                     K. Danzmann\\
               University of Hannover\\
\htmladdnormallink{kvd@mpqgrav2.amp.uni-hannover.de}
{mailto:kvd@mpqgrav2.amp.uni-hannover.de}\\
\end{center}

GEO600 is a laser interferometric gravitational wave detector with 600
m long arms being built in the small town of Ruthe, near Hannover,
Germany. It is designed and constructed by a British-German
Collaboration comprising the research groups from University of
Glasgow (Jim Hough), University of Cardiff and
Albert-Einstein-Institut (Bernard Schutz), and University of Hannover
and Max-Planck-Institut f\"ur Quantenoptik (Karsten Danzmann).

The objective is to use advanced technology right from the beginning and to
achieve a sensitivity not too far from first generation LIGO and VIRGO.
GEO600 will serve as a testbed for second generation detector concepts and
possibly take part in the first round of coincidence observations. GEO600 is
a somewhat smaller instrument, but is meant to be very flexible and can be
built on a short time-scale. Because the detector is not designed to be
extensible in length, the total capital cost of the project can be kept to
about 7 M\$. 

Groundbreaking for GEO600 was in September 1995. Due to an unusually cold 
winter, construction was delayed for several month. But this month the 
buildings and the trench for the submerged vacuum tube were finished.
The vacuum tube has a diameter of 60 cm and is of an unusual but 
cost-effective design that has been proposed by Roger Bennett from
Rutherford Appleton Laboratory. We are using a wall thickness of only
0.8 mm and the tube is stiffened by a continuous corrugation of about
1 inch amplitude that runs along the whole length of the tube. No bellows 
are thus required to take up the thermal expansion.
The tube is suspended inside the trench by a wire pendulum from
rollers running along a rail. The vacuum tube is manufactured in 4 m 
long segments that are delivered to the site, welded to the rest of 
the tube in the eastern end building and then pushed into the trench.
Welding and installation of the tube on the site are beginning
in the first week of September.

More information about GEO600 can found at our web site\\
\htmladdnormallink{\protect {\tt http://www.geo600.uni-hannover.de}}
{http://www.geo600.uni-hannover.de}
 
\vfill\eject 
\section*{\centerline {Update on Black Hole Microstates in String Theory}}
\addtocontents{toc}{\protect\medskip}
\addcontentsline{toc}{subsection}{\it Black hole microstates in string
theory by Gary Horowitz}
\begin{center}
\medskip
Gary T. Horowitz, UC Santa Barbara\\
\htmladdnormallink{gary@cosmic.physics.ucsb.edu}
{mailto:gary@cosmic.physics.ucsb.edu}
\end{center}

Last January, Strominger and Vafa \htmladdnormallink{(hep-th/9601029)}
{http://xxx.lanl.gov/abs/hep-th/9601029}  showed that the Bekenstein-Hawking
entropy of a static five dimensional extreme black hole was precisely 
reproduced by counting states in string theory with the same mass and charge
(for macroscopic black holes). This touched off an explosion of interest
and in the next few months, this agreement was shown to hold for near
extremal as well as extremal, four and five dimensional black holes, including
rotation. I wrote a review of these developments in April 
\htmladdnormallink{(gr-qc/9604051)}
{http://xxx.lanl.gov/abs/gr-qc/9604051}. What I would
like to do here is summarize some of the progress since then.

Perhaps the most important new development is a calculation by Das and
Mathur \htmladdnormallink{(hep-th/9606185)}
{http://xxx.lanl.gov/abs/hep-th/9606185} showing that the {\it rate}
of Hawking radiation from a near extremal black hole agrees with the
string theory prediction based on interactions between the
microstates. The fact that the spectrum is thermal with the same
temperature as the black hole is not a surprise, given that it was
already known that the entropy as a function of energy was the same in
the two systems. However, the fact that the overall coefficient agrees
is highly nontrivial and quite remarkable. This result has
implications for the black hole information puzzle. Recall that in
string theory, there is a length scale $l_s$ set by the string
tension.  Newton's constant is related to this length and the string
coupling $g$ by $G= g^2 l_s^2$ (in four dimensions).  At weak
coupling, $g \ll 1$, an extreme black hole is described by a flat
space configuration of objects known as D-branes. A near extremal
black hole is described by an `excited state' of D-branes. In this
description, there is no analog of the event horizon and the emission
from excited D-branes is manifestly unitary.  The apparent thermal
nature of the radiation arises from the large number of degress of
freedom, just like an ordinary hot object.  At strong coupling, the
gravitational field becomes stronger and one obtains a near extremal
black hole. The fact that the rate of Hawking evaporation from this
black hole agrees with the string calculation is further evidence that
radiation from near extremal black holes is also unitary.

In another development, there has been a great increase in the class
of solutions for which the Bekenstein-Hawking entropy has been shown
to agree with the counting of string states.  Previously, it was shown
that for black holes depending on a finite number of parameters
(including mass, charges and angular momentum) the entropy as a
function of these parameters was reproduced by counting states of
D-branes at weak string coupling. Recently with Don Marolf,
we extended this to 
the case where the solution depends on arbitrary {\it functions} 
(\htmladdnormallink{hep-th/9605224}{http://xxx.lanl.gov/abs/hep-th/9605224},
\htmladdnormallink{hep-th/9606113}{http://xxx.lanl.gov/abs/hep-th/9606113}).

One does not usually expect a solution with an event horizon to depend
on arbitrary functions, since the `no-hair' theorems show that
stationary black holes are characterized by only a few parameters. If
one tries to add a wave to the spacetime, it either falls down the
hole, or radiates to infinity. However it turns out that extremal
black strings, i.e. one dimensional extended objects with an event
horizon, are different (Larsen and Wilczek
\htmladdnormallink{hep-th/9511064}{http://xxx.lanl.gov/abs/hep-th/9511064}).
They can support traveling waves of arbitrary profile. These waves
affect the horizon area and the distribution of momentum along the
black string. By counting states of D-branes with the same momentum
distribution as the black string, one finds perfect agreement with the
Bekenstein Hawking entropy for all wave profiles
(\htmladdnormallink{hep-th/9605224}{http://xxx.lanl.gov/abs/hep-th/9605224},
\htmladdnormallink{hep-th/9606113}{http://xxx.lanl.gov/abs/hep-th/9606113}).

An outstanding open question is to
extend these results to black holes which are far from extremality. 
There are indications that we are getting close to taking this
important next step.

\vfill\eject
\section*{\centerline {LIGO project status}}
\addtocontents{toc}{\protect\medskip}
\addcontentsline{toc}{subsection}{\it LIGO project status by Stan Whitcomb}
\begin{center}
\medskip
Stan Whitcomb, Caltech \\
\htmladdnormallink{stan@ligo.caltech.edu}{mailto:stan@ligo.caltech.edu}
\end{center}

Construction continues to move forward rapidly at both LIGO sites
(Hanford Washington and Livingston, Louisiana).  At the Hanford site,
construction of 8 kilometers of concrete foundations which will support
the beam tubes has been completed.  The final survey of the foundation
along the two arms indicates that they are straight and level with an
accuracy of 1.5 cm.  Our Architect/Engineering contractor (Ralph M.
Parsons Co.) completed the final design for the buildings.  A contract
with Levernier Construction Inc of Spokane Washington for the building
construction was signed, and work is now underway.  At the Louisiana
site, the main activity is the rough grading (earthwork to level the
site and to build up a berm on which the LIGO facility will be built).
This work has gone more slowly than expected due to heavy rains, but is
now nearing completion.

The vacuum system is also moving forward. Chicago Bridge and Iron, the
company building the LIGO beam tubes (which connect the vertex and ends
of the two arms), is installing its fabrication equipment in a facility
near the Hanford site. They are preparing for full production of the
LIGO beam tubes and plan to begin installation by fall of this year.
The final design of the vacuum chambers and associated equipment which
will be in the located in the buildings has been completed.  Our
contractor for this effort, Process Systems International, is now
building the first large chambers.

The design of the LIGO detectors is accelerating, with most detector
subsystems well into the preliminary design phase.  Orders have already
been placed for the fused silica that will be used for the test masses
and other large optics.  LIGO's decision to switch its baseline
interferometer design to Nd:YAG lasers operating 1.06 microns has led
to a development contract with Lightwave Electronics Corporation to
develop a 10 W single frequency laser; first results from this
development are expected near the end of the year.

In the R\&D program, the 40 m interferometer has been converted to an
optically recombined system as the first step toward recycling.  The
signal extraction and control topology in the recombined configuration
is similar to that planned for the full-scale interferometers; a prime
objective of this effort was to compare these signals with the results
of modeling and in particular to study the problem of lock
acquisition.  At MIT, optical phase noise at the level of $10^{-10}$
rad Hz$^{-1/2}$ are being investigated with a 5 m long suspended
interferometer. This interferometer, initially configured as a simple
Michelson, has now been converted to a recycled configuration. The
increase in effective power due to recycling is approximately a factor
of 500, leading to nearly 100 W incident on the beamsplitter.  A
detailed characterization of the noise is presently underway.

As an additional means of communicating up-to-the-minute information
about LIGO, we have initiated a monthly newsletter.  It can be accessed
through our WWW home page at 
\htmladdnormallink{\protect {\tt http://www.ligo.caltech.edu}}
{http://www.ligo.caltech.edu}.

\vfill\eject 
\section*{\centerline {The Hamiltonian constraint}\\\centerline{in
       the loop representation of quantum gravity}}
\addtocontents{toc}{\protect\medskip}
\addcontentsline{toc}{subsection}{\it The Hamiltonian constraint of
quantum gravity and loops by John Baez}

\begin{center}
John Baez, UC Riverside\\
\htmladdnormallink{jbaez@math.ucr.edu}{mailto:jbaez@math.ucr.edu}
\end{center}

For some time now, the most important outstanding
problem in the loop representation of quantum gravity
has been to formulate the Wheeler-DeWitt equation in a
rigorous way by making the Hamiltonian constraint into a
well-defined operator.   Thomas Thiemann recently wrote
four papers aimed at solving this problem
(\htmladdnormallink{gr-qc/9606088}
{http://xxx.lanl.gov/abs/gr-qc/9606088},
\htmladdnormallink{89}
{http://xxx.lanl.gov/abs/gr-qc/9606089},
\htmladdnormallink{90}
{http://xxx.lanl.gov/abs/gr-qc/9606090},
\htmladdnormallink{91}
{http://xxx.lanl.gov/abs/gr-qc/9606091})
which have caused quite a bit
of excitement among those working on the loop representation.
In this brief introduction to his work and the history leading up to it,
I will not attempt to credit the many people to whose work I allude;
detailed references can be found in his papers.

An interesting feature of Thiemann's approach is that while it
uses the whole battery of new techniques developed in the loop
representation of quantum gravity, in some respects it returns to
earlier ideas from geometrodynamics.  Recall that in
geometrodynamics {\it \'a la} Wheeler and DeWitt, the basic
canonically conjugate variables were the 3-metric $q_{ab}$ and 
extrinsic curvature $K^{ab}$.   The idea was to quantize these,
making them into operators acting on wavefunctions on the space
of 3-metrics, and then to quantize the Hamiltonian and
diffeomorphism constraints and seek wavefunctions annihilated
by these quantized constraints.   In particular, if $H$ denotes the 
Hamiltonian constraint, a physical state $\psi$ should satisfy the
Wheeler-DeWitt equation 
\begin{equation}
          H \psi = 0 .
\end{equation}

However, this program soon became regarded as dauntingly
difficult for various reasons, one being that $H$ is not a
polynomial in $q_{ab}$ and $K^{ab}$: it contains a factor of
$({\det q})^{1/2}$.  Experience had taught field theorists that it
is difficult to quantize non-polynomial expressions in the
canonically conjugate variables.

In the 1980's Ashtekar found a new formulation of general
relativity in which the canonically conjugate variables are a
densitized complex triad field $E^a_i$ and a chiral spin
connection $A_a^i = \Gamma_a^i - iK_a^i$, where $\Gamma_a^i$ is
built from the Levi-Civita connection of the 3-metric and
$K_a^i$ is built from the extrinsic curvature.  As their names
suggest, $E^a_i$ and $A_a^i$ are analogous to the electric field
and vector potential in electromagnetism.

At first glance, in terms of $E^a_i$ and $A_a^i$ the Hamiltonian
constraint appears polynomial in form.  This greatly revived
optimism in canonical quantum gravity.   However, in this new
formalism one is really working with the {\it densitized} Hamiltonian
constraint $\tilde H$, which is related to the original
Hamiltonian constraint by $\tilde H = (\det q)^{1/2} H$.  Thus in
a sense the original problem has been displaced rather than
addressed.   It took a while, but it was eventually seen that
many of the problems with quantizing $\tilde H$ can be traced to
this fact (or technically speaking, the fact that it has density
weight 2).  

A more immediately evident problem was that because $E^a_i$ is
complex-valued, the corresponding 3-metric is also complex-valued
unless one imposes extra `reality conditions'.   The reality
conditions are easy to deal with in the Riemannian theory, where
the signature of spacetime is taken to be $++++$.  There one can
handle them by working with a {\it real} densitized triad field
$E_i^a$ and an ${\rm SU}(2)$ connection given by $A_a^i = \Gamma_a^i +
K_a^i$.  In the physically important Lorentzian theory, however,
no such easy remedy is available.

Despite these problems, the enthusiasm generated by the new
variables led to a burst of work on canonical quantum gravity. 
Many new ideas were developed, most prominently the loop
representation.  In the Riemannian theory, this gives a perfectly
rigorous way to construct the Hilbert space on which the
Hamiltonian constraint is supposed to be an operator: the Hilbert
space $L^2({\cal A})$ of square-integrable wavefunctions on the space
${\cal A}$ of ${\rm SU}(2)$ connections.  The idea is to work with graphs
embedded in space, and for each such graph to define a Hilbert
space of wavefunctions depending only on the holonomies of the
connection along the edges of the graph.   One then forms the
union of all these Hilbert spaces and completes it to obtain the
desired Hilbert space $L^2({\cal A})$.   

It turns out $L^2({\cal A})$ has a basis of `spin networks', given by
graphs with labellings of the edges by representations of
${\rm SU}(2)$ --- i.e., spins --- as well as certain labellings of
the vertices.  One can quantize various interesting observables
such as the area of a surface or the volume of a region of space,
obtaining operators on $L^2({\cal A})$.   Moreover, the matrix elements
of these operators have been explicitly computed in the spin
network basis.

Thiemann's approach applies this machinery to Lorentzian gravity
by exploiting the interplay between the Riemannian and Lorentzian
theories.  As in the Riemannian theory, he takes as his
canonically conjugate variables a real densitized triad field
$E_i^a$ and an ${\rm SU}(2)$ connection $A_a^i$.  This automatically
deals with the reality conditions.   He also takes as his Hilbert
space the space $L^2({\cal A})$ as defined above, since it turns out
that this space is acceptable for the Lorentzian theory as well
as the Riemannian theory.   Then, modulo some important
subtleties we discuss below, he quantizes the Hamiltonian
constraint of Lorentzian gravity to obtain an operator on
$L^2({\cal A})$.  Interestingly, it is crucial to his approach that he
quantizes $H$ rather than the densitized Hamiltonian constraint
$\tilde H$.  This avoids the regularization problems that plagued
attempts to quantize $\tilde H$.

How does Thiemann quantize the Hamiltonian constraint?  First, in
the context of classical general relativity he derives a very
clever formula for the Hamiltonian constraint in terms of the
Poisson brackets of the connection $A_a^i$, its curvature
$F_{ab}^i$ --- analogous to the magnetic field in electromagnetism
--- and the total volume $V$ of space.  (For simplicity, we assume
here that space is compact.)   Using the trick of replacing
Poisson brackets by commutators, this reduces the problem of
quantizing the Hamiltonian constraint to the problem of
quantizing $A_a^i$, $F_{ab}^i$, and $V$.  As noted, $V$ has already
been successfully quantized, and the resulting `volume operator'
is known quite explicitly.   This leaves $A_a^i$ and $F_{ab}^i$.

Now, a fundamental fact about the loop representation --- at
least as currently formulated --- is that the connection and
curvature do not correspond to well-defined operators on
$L^2({\cal A})$, even if one smears them with test functions in the
usual way.    Instead, one has operators corresponding to
parallel transport along paths in space.  Classically we can
write a formula for $A_a^i$ in terms of  parallel transport along
an infinitesimal open path, and a formula for $F_{ab}^i$ in terms
of parallel transport around an infinitesimal loop.  However, in
loop representation of
the quantum theory one cannot take the limit as the path or
loop shrinks to zero length.  The best one can do when quantizing
$A_a^i$ and $F_{ab}^i$ is to choose some paths or loops of finite
size and use parallel transport along them to define {\it
approximate} versions of these operators.  This introduces a new
kind of ambiguity when quantizing polynomial expressions in
$A_a^i$ and $F_{ab}^i$: dependence on arbitrary choices of
paths or loops.

So, contrary to the conventional wisdom of old, while the factors
of $(\det q)^{1/2}$ in the Hamiltonian constraint are essential
in Thiemann's approach, the polynomial expressions in $A_a^i$ and
$F_{ab}^i$ introduce problematic ambiguities!  In short, Thiemann
really constructs a large family of {\it different versions} of
the Hamiltonian constraint operator, depending on how the choices
of paths and loops are made.  However, by making these choices
according to a careful method developed with the help of Jerzy
Lewandowski, the ambiguity is such that two different versions
acting on a spin network give spin networks differing only by a
diffeomorphism of space.  Mathematically speaking we may describe
this as follows.   Let $L \subset L^2({\cal A})$ be the space of finite
linear combinations of spin networks, and let $L/{\rm Diff}$ be the
space of finite linear combinations of spin networks modulo
diffeomorphisms. Then Thiemann obtains, for any choice of lapse
function $N$, a smeared Hamiltonian constraint operator  
\begin{equation}
      \hat H(N) \colon L \to L/{\rm Diff}  ,
\end{equation}
independent of the arbitrary choices he needed in his construction.

Since these operators $\hat H(N)$ do not map a space to itself we
cannot ask whether they satisfy the naively expected commutation
relations, the `Dirac algebra'.   However, this should come as no
surprise, since the Dirac algebra also involves other operators
that are ill-defined in the loop representation, such as the
3-metric $q_{ab}$.  Thiemann does check as far as possible that
the consequences one would expect from the Dirac algebra really
do hold.  Thus if one is troubled by how arbitrary choices of
paths and loops prevent one from achieving a representation of
the Dirac algebra, one is really troubled by the assumption,
built into the loop representation, that $q_{ab}$, $A_a^i$, and
$F_{ab}^i$ are not well-defined operator-valued distributions. 
Ultimately, the validity of this assumption can only be known
through its implications for physics.  

Thiemann's approach to quantizing the Hamiltonian constraint is
certainly not the only one imaginable within the general
framework of the loop representation.  (Indeed, his papers
actually treat two approaches, one yielding a formally Hermitian
operator, the other not.)  As soon as his work became understood,
discussion began on whether it gives the right
physics, or perhaps needs some modification, or perhaps exhibits
fundamental problems with the loop representation.  The quest for a
good theory of quantum gravity is far from over.  But at the very
least, Thiemann's work overturns some established wisdom and
opens up exciting new avenues for research.

\eject 
\section*{\centerline{International conference on gravitational waves:}\\
\centerline{Sources and Detectors}}
\addtocontents{toc}{\protect\bigskip}
\addtocontents{toc}{\bf Conference reports:}
\addtocontents{toc}{\protect\medskip}
\addcontentsline{toc}{subsubsection}{\it International conference on
gravitational waves by Valeria Ferrari}
\begin{center}
\medskip
Valeria Ferrari and Maria Alessandra Papa,
 Universit\` a di Roma\\
\htmladdnormallink{valeria@roma1.infn.it}{mailto:valeria@roma1.infn.it}
\end{center}

The Conference was held on March 19-23 1996 in Cascina (Pisa) near the
site where the VIRGO interferometer is now under construction. It was
attended by 120 physicists plus a sociologist who is ``keeping under
observation" the scientific community involved in the search of
gravitational waves.  The aim of the Conference was to gather the
efforts of the theoreticians and the experimentalists working in the
field and stimulate future work on the phenomenology of GWs in close
connection with the experiments.

On the theoretical side, the sources of GWs have been the
subject of several talks.
Inspiralling compact binaries have been discussed by L.
Blanchet, who showed that,  in order to extract significant
information from VIRGO and LIGO observations, the radiation
field and the internal dynamics of the binary system
must be evaluated including post-newtonian corrections at
least up to third order.
E. Gorgoulhon and S. Bonazzola have discussed how efficiently
a magnetic dipole moment of a rotating neutron star can
induce  distortions in the axial symmetry with consequent 
emission of GWs. Other mechanisms which may be responsible for
axial symmetry breaking (such us Chandrasekhar-Friedman-Schutz
instability, MacLaurin-Jacobi transition
and crust defects) have also been reviewed.
K. Kokkotas has shown  that from the detailed knowledge of the
spectrum of the quasi-normal modes of a compact star
one can infer  the values of its mass and radius 
and have indications on its internal structure.
Great interest have received the  
estimates of the relic stochastic GW background spectrum
provided by inflationary cosmology in the framework of string
theory, which have been presented by G.Veneziano and R. Brustein.
They suggest that, depending on the constraints of the theory,
the predicted power spectra may be detectable.
Another kind of stochastic GW background
due to cosmological supernovae explosion, has been evaluated
(D. Blair \ $z\geq 2,$ and A. Di Fazio-V.Ferrari \ $4 \leq z \leq 8$),
and it emerges that it may be competitive with the 
string background in the VIRGO-LIGO bandwidth.

Fully relativistic numerical codes for gravitational collapse
and coalescing compact objects have been shown to be in progress.

The status of the experiments was discussed both in plenary
talks and workshops.
A number of resonant bars are actually taking data
as well as the  TENKO-100 interferometer in Japan.
The quoted sensitivities to a pulse
of GWs for the resonant experiments are:

\noindent EXPLORER (Geneva, Cern) $ h\sim 6\cdot 10^{-19},$  \\
NAUTILUS (Frascati LNF, Roma), $ h\sim 6\cdot 10^{-19},$
\\
NIOBE (Perth, UWA) $ h\sim 6\cdot 10^{-19}$, \\
TENKO-100 (ISAS Japan) 
$\tilde h\sim 10^{-19} \frac{1}{\sqrt{Hz}}@100 Hz,$
and $\tilde h\sim 5\cdot 10^{-19}
\frac{1}{\sqrt{Hz}}@1kHz$.\\
AURIGA (Legnaro LNL, Padova) started the cryogenic tests
and will  soon be operational.

The state of the art for
the interferometric antennas,  VIRGO, LIGO, GEO600, and TAMA,
has been reported and  the following expected sensitivities
have been quoted:

\noindent VIRGO: $\tilde h\sim 10^{-21} \frac{1}{\sqrt{Hz}}@ 10
Hz,$ and
$ \tilde h\sim 3\cdot 10^{-23} \frac{1}{\sqrt{Hz}}@ 500
Hz.$\\
LIGO: $ \tilde h\sim 2\cdot 10^{-23} \frac{1}{\sqrt{Hz}}$
in a bandwidth of $ \sim 200 Hz.$\\
GEO600: $ \tilde h\sim 4\cdot 10^{-23} \frac{1}{\sqrt{Hz}}$
(depending on bandwidth)\\
TAMA: $ \tilde h\sim 8\cdot 10^{-23} \frac{1}{\sqrt{Hz}}
@300 Hz$
in a bandwidth of $ \sim 300 Hz.$

Doppler tracking experiments and upper limits on the emission
of GWs
in the range of $ 10^{-4}-10^{-1} Hz$ have been reviewed
by Luciano Iess.
The sensitivity  of LISA to GWs from various binary systems
(WD-WD, BH-BH, WD-BH, MBH-MBH etc)   and to GWs of
cosmological origin, has been discussed, together with the
planning of the experiment, by Peter Bender.
{}From his graphs  LISA's sensitivity should range between
$ \tilde h\sim 10^{-21} \frac{1}{\sqrt{Hz}}@ 10^{-4} Hz,$
and $ \tilde h\sim 10^{-23}\frac{1}{\sqrt{Hz}}@ 10^{-1}Hz.$

Data analysis for extracting GW-signals from present and future
data, has focused essentially on the study of filtering procedures
for single detectors and for different kinds of networks.
These two issues have been discussed in
talks regarding the use of  APE1000
to detect coalescing binaries and pulsars parameters
(A.Vicere'), 
the search of monocromatic and stochastic GWs with NAUTILUS
and EXPLORER (P.Astone), the estimate of chirp parameters
(I.M.Pinto), 
the signal deconvolution for a multimode spherical
detector (E.Coccia),
the cross-correlation of data from several bars
(S. Vitale),  the use of bar-interferometer
networks  for pulse detection (B. Schutz), and the use
of local arrays of 
small resonators for high frequency detection (S.Frasca).

\eject
\section*{\centerline {12th Pacific Coast Gravity Meeting - 
Karel Kucha\v{r} fest}}
\addtocontents{toc}{\protect\medskip}
\addcontentsline{toc}{subsubsection}{\it PCGM12/KKfest by Richard
Price}
\begin{center}
Richard Price, University of Utah\\
\htmladdnormallink{rprice@mail.physics.utah.edu}
{mailto:rprice@mail.physics.utah.edu}
\end{center}

Near the end of March, at the University of Utah, there were two
relativity meetings that were loosely associated, at least in timing,
and which made for an interesting juxtaposition. On Thursday, March 21
was``KKfest," a one day conference honoring the 60th birthday of Karel
Kucha\v{r}. It was followed by the two days of the twelfth Pacific
Coast Gravity Meeting. The latter is a meeting centered on young
people; all talks are contributed, and each speaker, first year
student or Nobel laureate, gets 15 minutes.  The KKfest, by contrast,
consisted of six invited talks, by``the establishment."  A banquet on
Friday evening honored Karel Kucha\v{r}, but was attended by almost
all the PCGM12 participants. Almost 100 people attended! And the
crossover was not limited to the banquet. Almost all participants in
each conference attended the other conference. It gave the venerable
sages of the KKfest a chance to be energized by the enthusiasm of
those starting out in the field; at the KKfest the young people of
PCGM12 got a first hand contact with some of the history of the ideas
in our field.

The speakers during the day of KKfest were Jiri Bi\v{c}\'{a}k, Bryce
DeWitt, Petr H\'{a}j\'{\i}\v{c}ek, Jim Hartle, Claudio Teitelboim, and
Jim York. All their talks gave a historical perspective on modern
issues, and on the influence on Karel Kucha\v{r}'s contributions. Talks in
the KKfest covered some exact solutions and black hole thermodynamics,
but the main focus, of course, was quantum gravity. Here reviews were
given and recent ideas were reported in the canonical approach, the
covariant approach, and generalized quantum mechanics.

The Pacific Coast Gravity meeting had 54 talks (!) by presenters from
22 institutions. (The Pacific coast was analytically extended to
include, for example, Ireland.)  The breadth of the topics showed the
recent breadth of our field. There were, on the one hand, talks on knot
polynomials (Jorge Pullin) and intermediate topologies (Don
Marolf). On the other there were reports on the low frequency
satellite tracking gravitational wave experiment (John Armstrong), and
on light baffles for the LIGO beam tube (Kip Thorne).

As in the past, Doug Eardley donated a prize to be awarded for the
best graduate student presentation. When given no choice but to point
to a single name, an impartial international jury pointed to the name
Shawn Kolitch of UC Santa Barbara.

Any short list of the most interesting presentations at PCGM would be
incomplete, but would include a reversal of a recent result, and a
verification of a longstanding one. Gary Horowitz (UCSB) reported
computations of black hole entropy from string theory. Previously such
calculations had been claimed to imply that extreme black holes had
zero entropy.  The correction of a technical error in those
calculations has led to new results which show that entropy for
extreme holes is related to horizon area exactly the same as for
moderate holes. Paul Anderson (Wake Forest) reported on a careful
study of a gravitational geon. His results completely confirmed the
claims in the classical paper by Brill and Hartle. Another talk that
stimulated much buzzing in the hallways was the claim by Thomas
Thiemann (Harvard) that a finite theory results if a real connection
is used for the Ashtekar variables.

At the Friday evening banquet the key speaker was John A. Wheeler who
applauded Karel Kucha\v{r}'s contributions, character and culture and
read some of the words of Vaclav Havel about the nature of our pursuit
of answers. A gentle roast followed and was enjoyed by all, or perhaps
by all but one.

\eject
\section*{\centerline {First International LISA Symposium}}
\addtocontents{toc}{\protect\medskip}
\addcontentsline{toc}{subsubsection}{\it First International LISA Symposium
by Robin Stebbins}
\begin{center}
Robin Stebbins, JILA/University of Colorado\\
\htmladdnormallink{stebbins@jila.colorado.edu}
{mailto:stebbins@jila.colorado.edu}
\end{center}

The First International LISA Symposium was held at the Rutherford Appleton
Laboratory in Chilton, 9-12 July 1996.  The symposium highlighted the
scientific opportunities of gravitational wave detection in space.  The
symposium was further enriched by poster sessions, technology
demonstrations, a full-scale mockup of a LISA spacecraft, laboratory tours
and a delightful dinner cruise on the Thames with live jazz!  The main oral
sessions are summarized below.  Selected papers from the symposium are
scheduled to appear in the March 1997 issue of Classical and Quantum
Gravity.  Mike Sandford, the scientific and local organizing committees, and
the RAL staff are to be commended for putting together such a stimulating
and pleasant symposium.

In the overview session, Rudiger Reinhard (ESA) described the status of LISA
in ESA's Horizons 2000 Plus Programme, and Karsten Danzmann (Hannover), Bill
Folkner (JPL) and Koos Cornelisse (ESTEC) described the current baseline
LISA mission.  Kip Thorne (Caltech) described a menagerie of dark, extremely
relativistic objects in the Universe which might be discovered with a
low-frequency gravitational wave detector in space, and the insight into
gravitation theories to be gained from them.  Martin Rees (Cambridge)
surveyed the available information on massive black holes and gave a very
positive assessment of the likelihood of detection of signals from various
scenarios.

The sources session focused on astrophysical systems which could produce
low-frequency gravitational waves likely to be detected by LISA.  Frank
Verbunt (Utrecht) reviewed the state of observational knowledge about
binaries systems consisting of main sequence stars and/or compact objects
which could give rise to detectable signals.  Steinn Sigurdsson (Cambridge)
described the capture of low-mass black holes by massive black holes in
galactic cusps.  Alberto Vecchio (MPI/Potsdam) reported on signals from
coalescing massive black holes.  Curt Cutler (Penn State) showed that LISA
could only identify the source of signals from coalescing massive black
holes if there was some supplementary optical indication.  Pete Bender
(JILA) described a revised estimation of the confusions limit from galactic
and extragalactic binaries.  I. Pinto (Salerno) described the spectrum of
signals from insular clusters.

The session on gravitational theories and numerical relativity began with a
talk by Richard Matzner (Texas) on the computation of waveforms from the
coalescence of black hole binary systems.  Leonid Grishchuk (Cardiff)
offered an explanation of cosmic background anisotropies based on relic
gravitational waves, and noted existing observational support.  Ewald Muller
(MPI/Garching) described the gravitational wave generation in the inner core
and the outer convective region of a type II supernova.  Gerhard Shafer
(MPI/Jena) discussed how alternate theories of gravitation might be checked
with LISA.  

Updates were given by on the VIRGO Project by Francesco Fidecaro (Pisa), on
GEO 600 by Harald Luck (MPI/Hannover), on LIGO by David Shoemaker (MIT), on
the TAMA Project by Keita Kawabe (Tokyo) experiments.  Construction  is
proceeding well on all of these ground-based interferometers.  Bruno
Bertotti (Pavia) reviewed past attempts to detect gravitational waves by
spacecraft tracking and previewed plans for the Cassini mission.  Guido
Pizzella (Rome) summarized the current and expected performance of resonant
detectors.  Stefano Vitale (Trento) analyzed the sensitivity of two resonant
detectors, and two resonant detectors and an interferometer, to an
isotropic, stochastic background.  M. Cerdonio (Padova) reported performance
of AURIGA when cooled to 140 mK.  

The next session delved into gravitational wave signal extraction and data
analysis.  Robin Stebbins (JILA) outlined the challenges of extracting
astrophysical information from the many and varied gravitational wave
signals likely to be in LISA data.  Michael Peterseim (Hannover) examined
the angular resolution obtainable with various signal parameters.  Roland
Schilling (MPI/Garching) analyzed the response function of LISA above 10 mHz
where the wavelength is shorter than the armlength.  Giacomo Giampieri (QMC)
discussed the anisotropy of the stochastic background caused by galactic
binaries, as seen by LISA.  Oliver Jennrich (Hannover) reported on the
polarization resolution which LISA could obtain.  Bill Folkner (JPL)
described the onboard signal processing planned for in the LISA mission.  L.
Milano (INFN) simulated the application of matched filters to search for
binary signals in VIRGO data.  

The final session of the conference addressed enabling technologies for
gravitational wave detection.  Sheila Rowan (Glasgow) reported on the
performance of prototype monolithic fused quartz suspensions for
ground-based interferometers.  Paul McNamara (Glasgow) described a
laboratory demonstration of weak light phase-locking, a requirement for
LISA.  Dan DeBra (Stanford) reviewed drag-free satellite technology, both
flown and future.  M. Rodrigues (ONERA) explained the LISA accelerometer
design.  Clive Speake (Birmingham) analyzed two designs for capacitive
sensing circuits for the dominant noise source, and showed that LISA goals
can be achieved with either design.  Yusuf Jafry (ESA) reported on a
simulation of cosmic ray charging of the LISA proof mass done with the GEANT
code.  And P. Rottengatter (LZH) reported on the successful intensity and
amplitude stabilization of Nd:YAG lasers for use in LISA.  Dave Robertson
(Glasgow) described the LISA optics and limiting noise sources in the
optical measurement.  Walter Winkler (MPI/Garching) gave an analysis of the
far-field effects of LISA transmitting a truncated Gaussian beam.  Wei-Tou
Ni (Taiwan) described the ASTROD mission for performing several relativistic
tests in solar orbit, and technology development activities to support
fundamental physics missions.  Dan DeBra read a paper submitted by S.
Marcuccio (Centrospacio) reporting on recent tests and development of the
Field Emission Electric Propulsion (FEEP) thrusters planned for LISA.

\eject
\section*{\centerline{Schr\"odinger Institute Workshop on}\\
\centerline{Mathematical Problems of Quantum Gravity}}
\addtocontents{toc}{\protect\medskip}
\addcontentsline{toc}{subsubsection}{\it Schr\"odinger Institute Workshop by
Abhay Ashtekar}
\begin{center}
Abhay Ashtekar, Penn State\\
\htmladdnormallink{ashtekar@phys.psu.edu}{mailto:ashtekar@phys.psu.edu}
\end{center}

A 2-month workshop was held at the Erwin Schr\"odinger International
Institute for Mathematical Sciences in Vienna during July and August,
'96. It was jointly organized by Peter Aichelburg and myself.

There were 23 participants from outside Austria, mostly young
physicists who have been working on various aspects of quantum
gravity. In addition, about a dozen faculty and students from Vienna
actively participated in the seminars and discussions. While the focus
of this effort was on non-perturbative quantum general relativity,
there were several experts from string theory, supergravity, quantum
cosmology, quantum field theory, as well as mathematical physics in a
broad sense of the term. Unfortunately, there was a rather severe
desk-space limitation in July and so the workshop had to make do
without the participation of a number of experts who had
time-constraints of their own. There were two weekly ``official
seminars" which were widely announced --one entitled ``fundamental
issues", and the other ``advanced topics". They enhanced the scientific
interaction between workshop participants and the local physics and
mathematics community. In addition, there were ``discussion seminars"
(the remaining) three days a week. The afternoons were left open
for further informal discussions (and real work!).

On the scientific front, the workshop elevated the subject to a new
level of maturity. It enabled the participants to take stock of a
number of areas to obtain a global picture of issues that are now
well-understood and also opened new directions for several other key
issues. Because of the space limitation, I will restrict myself here
only to a few illustrative highlights. A more detailed discussion of
the (July) activities can be found in John Baez's ``This Week's Finds"
series, weeks 85-88 (
\htmladdnormallink{\protect {\tt http://math.ucr.edu/home/baez/twf.html}}
{http://math.ucr.edu/home/baez/twf.html}) which
also contains many references. A Schr\"odinger Institute pre-print
containing abstracts of seminars will be available early
October. Further information on the workshop as well as pre-prints of
research carried out during the workshop can be obtained from the
Schr\"odinger Institute home page
\htmladdnormallink{\protect {\tt http://www.esi.ac.at/ESI-Preprints.html}}
{http://www.esi.ac.at/ESI-Preprints.html}).

In the list that follows, the names in parenthesis refer to people who
gave seminars or led discussions (although almost everyone present
made significant contributions to all the discussions).

\medskip

\noindent{\sl Quantum Hamiltonian constraint.} (Hans-J\"urgen Matschull,
Jorge Pullin, Carlo Rovelli, Thomas Thiemann)

\noindent{\sl Quantum geometry.} (AA, Jerzy Lewandowksi, Renate Loll, 
Thiemann)

\noindent{\sl Lattice methods and skeletonization in loop quantum gravity.}
(Loll, Michael Reisenberger)

\noindent{\sl Super-selection rules in quantum gravity.} (AA, Lewandowski,
Donald Marolf, Jose Mour\~ao, Thiemann)

\medskip\goodbreak

\noindent{\sl Degenerate metrics: extensions of GR.} (Ted Jacobson,
Lewandowski, Matschull)

\noindent{\sl Global issues, Hamiltonian formulations.} (Fernanado Barbero,
Domenico Giulini)

\noindent{\sl Mathematical issues in quantum field theory and quantum
gravity.}  (John Baez, Matthias Blau, Herbert Balasin, Rodolfo
Gambini, Mourao, Marolf)

\noindent{\sl Exactly soluble midisuperspaces.} (AA, Hermann Nicolai)

\noindent{\sl Lessons from low dimensional gravity.} (AA, Giulini,
Lewandowski, Marolf, Mourao, Thiemann, Strobl).

\medskip

\noindent{\sl Black-hole entropy.} (Jacobson, Kirill Krasnov, Marolf,
Rob Myers, Rovelli)

\noindent{\sl Topological quantum field theories} (Baez, Reisenberger)

\noindent{\sl String duality, conformal field theories} (J\"urgen Fuchs,
Krzysztof Meissner, Myers, Strobl)

\noindent{\sl Foundations of quantum mechanics and quantum cosmology}
(AA, Giulini, Jonathan Halliwell, Franz Embacher)

\medskip

If participants were to single out one topic that generated most
excitement, it would probably be the regularization of the Hamiltonian
constraint by Thiemann
(\htmladdnormallink{gr-qc/9606088}
{http://xxx.lanl.gov/abs/gr-qc/9606088},
\htmladdnormallink{89}
{http://xxx.lanl.gov/abs/gr-qc/9606089},
\htmladdnormallink{90}
{http://xxx.lanl.gov/abs/gr-qc/9606090},
\htmladdnormallink{91}
{http://xxx.lanl.gov/abs/gr-qc/9606091}). This has significantly deepened our
understanding of the mathematical problems underlying quantum dynamics
of general relativity. (For details, see Baez's article in this
issue.)  However, a number of important problems remain.  In
particular, during the workshop it was realized that these regularized
quantum constraints have the feature that they strongly commute not
only on diffeomorphism invariant states (which is to be expected
physically) but also on a rather large class of states which are not
diffeomorphism invariant (which is alarming from a physical
viewpoint). A related potential difficulty is with the semi-classical
limit: it is not clear if all the quantum constraints, taken together,
admit a sufficient number of semi-classical states.  Analogous
calculations in 2+1 dimensions indicate that the appropriate
semi-classical sector {\it does} exist.  In 3+1 dimensions, further
work is needed. This will no doubt be an area of much research and new
effort in the coming year.

\eject
\section*{\centerline{Relativistic Astrophysics: a summer school 
at Bad Honnef}}
\addtocontents{toc}{\protect\medskip}
\addcontentsline{toc}{subsubsection}{\it Relativistic Astrophysics at
Bad Honnef by Hans-Peter Nollert}
\begin{center}
Hans-Peter Nollert, Penn State\\
\htmladdnormallink{nollert@phys.psu.edu}{mailto:nollert@phys.psu.edu}
\end{center}

Modern astrophysics is unthinkable without the input of general
relativity. Therefore, the German Astronomical Society (Astronomische
Gesellschaft) joined forces with the `Gravitation and Relativity
Theory' section of the German Physical Society (DPG) in organizing
this school on selected topics in relativistic astrophysics, such as
gravitational lensing, gravitational waves, neutron stars and
collapsing binaries, and accretion phenomena. The school took place in
the physics center of the Deutsche Physikalische Gesellschaft in Bad
Honnef from  August 19 to 23 1996.

{\bf J\"urgen Ehlers} brought the lectures off to a great start with his
comprehensive overview over the basic concepts of general relativity,
with emphasis on physical interpretation, on astrophysical relevance,
especially for lensing and gravitational radiation, and on the initial
value problem for time evolution calculations. He found ways to help
even old experts in the field see many things in a new light.

{\bf Peter Schneider} discussed gravitational lensing: its history in
the context of astrophysics, the basic concepts and the wealth of
information that can be gained from observations of weak lensing: Mass
profiles of galaxies, dark mass concentration, mean distribution of
galaxies, even the Hubble constant - and much more. With new
telescopes soon becoming operational, he foresees a bright future for
his field.

{\bf Joachim Wambsganss} described the searches for microlensing
events. He presented the theoretical background and an overview over
the history of the MACHO, EROS, and OGLE projects. The search for dark
matter objects, for binaries and planets is the main objective in
studying galactic microlensing events. About three times as many
events as expected are observed in the galactic bulge, but fewer than
expected towards the large Magellanic cloud. A preliminary conclusion
states that the galactic halo almost certainly does not consist of
brown dwarfs. The focus of attention for extragalactic events is on
the determination of size and brightness profile of the sources, and
on the detection of compact objects for the lenses and the
determination of their masses.

{\bf Ute Kraus} discussed theory and consequences of light deflection
near neutron stars. Geometric effects, such as increased visibility of
the star's surface, can have drastic effects for the pulse profiles of
radiation emitted on or near the surface of the star. Since her main
concern were light curves of X-ray pulsars, it is sufficient to
consider photon trajectories in a Schwarzschild metric. In addition to
the geometric effects, changes of photon energy and intensity
radiation have to be taken into account.

{\bf Karsten Danzmann's} guide on ``How to build a GEO600 interferometric
gravitational wave detector in your back yard with spare change found
under your couch cushions'' covered every aspect from using recycling
to make your laser light go further, to reducing noise of nearby
tractors, to the proper way of welding the vacuum tubes. If you can
spare a little more, go for LISA, the heavenly version - you'll be
first on the block to have one, and you will be guaranteed a variety
of spectacular sources, such as coalescence of massive black holes
anywhere in the universe, or white dwarf binaries.

{\bf Ed Seidel} reported that the Grand Challenge community is getting
ready to tackle a new challenge: the fully relativistic,
three dimensional treatment of the merger of neutron star binaries. The
plan is to use post-Newtonian techniques for the pre-coalescence
phase, and then take the results of this as initial data (at a
separation of about $8M$) for the general relativistic hydrodynamical
calculation. The relativistic field equations will be even more
difficult to handle than the hydrodynamic equations, requiring the
development of suitable algorithms, of adaptive mesh techniques,
finding the best gauge conditions, and an effective use of parallel
algorithms. 

{\bf Heinz Herold} discussed the effects of various equations of state
and of rapid rotation on the equilibrium state of neutron stars. The
structure equations can be solved using a variational principle in the
form of a minimum surface problem. The numerical treatment is based on
a finite element discretization. It turns out that higher mass models
allow higher angular velocities. The deformation of the surface of the
star was visualized using isometric embedding (for its internal
geometry) or ray-tracing (for a view from the outside). 

An excursion to the Drachenfels, a nearby hill featuring ancient ruins
of a fortress, a grand view over the Rhine river, and a restaurant,
provided some welcome diversion for the participants on Tuesday
afternoon. 

Instabilities of rotating stars can be quite frightening: {\bf Lee
Lindblom} pointed out that in principle, every star, even the earth,
shows rotational instability. Luckily, they are usually countered by
dissipative effects. Using a Newtonian two-potential technique, he
found that the balance may be in favor for the instabilities in the
case of neutron stars. However, it is not clear if they can prevail in
a relativistic context, since they will be damped by gravitational
radiation. At least for realistic equations of state, they may turn
out not to be an issue.

{\bf Hans-Peter Nollert} discussed treating collisions of black holes
and neutron stars without the help of supercomputers. He pretended
that the two colliding bodies are like a perturbation of the single
final object: $1 + 1 \approx 1$. The gravitational radiation emitted
during and after the collision can then be obtained from linear
equations. For black holes, the comparison with the full numerical
calculations is remarkably good. He wishes he could do the same trick
for neutron stars - if only a good fairy would take care of the
initial conditions...

Whatever the central source of a gamma ray burster actually is, there
has to be a fireball - unless gamma ray bursters are local, i.e. less
than 200pc away. With this premise, {\bf Peter M\'esz\'aros} gave
theoretical explanations for many observed features of these elusive
objects, based on the expansion of the fireball and the dissipation of
its energy. 

{\bf Harald Riffert} provided the necessary ingredients for a model of
thin accretion disks around black holes: Solve the gas dynamics in the
equatorial plane of a Kerr background metric, build the
energy-momentum tensor from an ideal fluid, viscous stress, and
radiation flux, assume the disk to be stationary and rotationally
symmetric, with velocities dominant in the $\phi$
direction. Integrating over the height of the disk, the vertical
structure equations decouple from the radial part. The radial disk
structure can be solved analytically, and the vertical equations have
the same form as in the Newtonian case. The resulting model spectra
can be fit to the UV-soft X-ray continuum of AGN.

When it comes to rapidly rotating relativistic systems, most work has
concentrated on black holes and neutron stars, with little attention
payed to other systems. {\bf Jim Ipser} studied rapidly rotating
accretion disks around compact objects, using a quasinormal mode
analysis for perturbations of a simple equilibrium model. As a clever
trick, he uses the perturbed Euler equation to eliminate the velocity
perturbations. Relying on the Cowling approximation eliminates the
metric perturbations, resulting in a single equation for a
potential-like fluid variable. Taking into account frame dragging, his
model provides a possible source for quasi-periodic oscillations in
black hole X-ray binaries, giving a counter-argument to the objection
that sources showing QPO's cannot be candidates for black holes.

Are quasars supermassive black holes or star clusters at the centers
of galaxies? After reviewing an impressive collection of observational
data, {\bf Max Camenzind} favored the black hole scenario.  In order
to explain the central machine providing the power and accelerating
the observed jets, magnetic fields are required. Consequently, the
magnetohydrodynamics of disks in the background field of rapidly
rotating stars was the topic of the second part of his lecture.

{\bf Fred Rasio} presented a three dimensional Newtonian treatment of the
merger phase of binary neutron star coalescence, using smoothed
particle hydrodynamics. A Newtonian treatment is interesting in its
own right: The hydrodynamics contain enough challenging physics, and
they dominate the dynamics of the merger. The results can thus serve
as preliminary estimates for the gravitational radiation emitted
during the merger. When fully numerical codes become available, the
Newtonian results can serve as a test case. In the future, nuclear
physics, strong relativistic effects, and turbulent viscosity should
be included for a more realistic treatment.

{\bf Pablo Laguna} studied the evolution of matter in curved spacetime,
using a smoothed particle approach on a fixed relativistic
background. The SPH simulation reproduces the results of length scale
estimates if the artificial viscosity is suitably adjusted. In
particular, he examined the tidal disruption of stars by massive black
holes. This scenario can be regarded as the fuelling process of active
galactic nuclei: A dense star cluster in the vicinity of a central,
supermassive black hole provides the necessary raw material.

A cosmic perspective was provided by {\bf Andreas Tammann}, who
reviewed observations determining the Hubble constant. Since
measurements of redshifts are generally undisputed, most of his talk
concentrated on determining cosmic distances. He used SNe Ia
supernovae calibrated by cepheids, the Virgo cluster, and field
galaxies. Including independent methods such as growth of supernovae
shells, gravitational lenses, or fluctuations of the microwave
background, he arrived at a value of $H_0 = 55 \pm 10$. He warned the
audience to be critical of headlines which will soon appear in popular
newspapers, claiming that a new distance determination of the Fornax
cluster in the southern hemisphere leads to $H_0 = 73$, since this
value may be based on improper identification of distances. He
discussed estimates for the age of the universe, which he puts at $12.5
-- 15 \times 10^9$ yrs, compatible with his favored value of $H_0$.

{\bf Michael Soffel} reviewed experiments relating to gravity: Is there
a fifth force (pronounced dead), does the gravitational constant
depend on time (not to within one part in $10^{11}$), and what is its
numerical value (the worst known physical constant)? He discussed the
weak equivalence principle, the Einstein EP, and the strong EP. All
are very well confirmed by various experiment; improved measurements
are desirable with respect to some quantizations of gravity, which
might cause tiny deviations ($10^{-11} -- 10^{-15}$). With regard to
general relativity, he discussed perihelion advance, light deflection,
timing delay, and the Lense-Thirring effect.

The proceedings of the school will be published by Vieweg in early
1997. 

The organizers, Hanns Ruder, Harald Riffert, and Hans-Peter Nollert
for the Astronomische Gesellschaft and Friedrich Hehl for the Deutsche
Physikalische Gesellschaft, wish to acknowledge the generous financial
support from the WE-Heraeus Foundation which made this school
possible.

We wish to point out that the names of the organizers of last year's
school on ``Relativity and scientific computing'', Friedrich Hehl and
Roland Puntigam for the Deutsche Physikalische Gesellschaft and Hanns
Ruder for the Astronomische Gesellschaft, were inadvertently left out
of the report on this school in the last issue of Matters of Gravity.

\eject
\section*{\centerline {Intermediate binary black hole workshop at Caltech}}
\addtocontents{toc}{\protect\medskip}
\addcontentsline{toc}{subsubsection}{\it Intermediate binary black hole
workshop by Sam Finn}
\begin{center}
Lee Samuel Finn, Northwestern\\
\htmladdnormallink{lsf@marlowe.astro.nwu.edu}
{mailto:lsf@marlowe.astro.nwu.edu}
\end{center}
Since late 1993, a wide collaboration of relativists have been engaged
in an effort to solve numerically for the final inspiral and
coalescence of a binary black hole system.  A quantitative
understanding of black hole binary coalescence is needed to complete
our solution of the relativistic Kepler problem, whose beginnings (in
a nearly Newtonian binary undergoing slow, adiabatic inspiral) and
endings (in a quiescent, single Kerr black hole) are already
understood separately.  The gravitational radiation arising from this
final stage of binary inspiral/coalescence may also be detectable in
the interferometric detectors now under construction; thus, the
waveforms predicted by these calculations may play an important role
in the associated data analysis.
 
To connect the initial and final states of the relativistic Kepler 
problem, or to use the predicted waveforms to learn something of the 
character of an observed coalescing binary, it is necessary that the 
initial data for the numerical calculation be firmly related to a 
binary system involving two distinct black holes of definite mass and 
spin in an orbit of certain energy and orbital angular momentum.  
Herein lies two problems:
\begin{enumerate}
\item Numerical calculations of coalescence so tax the anticipated 
computing resources expected to be available with next generation 
supercomputers that the numerical initial data must be imposed no 
earlier than $\sim4\pi$ orbital phase before coalescence.  At this 
separation the binary systemÕs total mass cannot be resolved into the 
individual black hole masses, nor can the systemÕs total angular 
momentum be usefully resolved into black hole spinÕs and orbital 
angular momentum.
\item The techniques used to evolve a binary from large separation, 
where its character (component masses and spins, orbital energy and 
angular momentum) can be described in Newtonian terms, to small 
separations, where the fully numerical evolution can begin, become 
increasingly suspect as the separation decreases; thus, either 
extensions to existing methods or entirely new methods must be found 
to continue the evolution of a binary system to the point where fully 
numerical methods can take over.
\end{enumerate}

To highlight the urgency of these problems, the Binary Black Hole 
Grand Challenge Alliance sponsored a one-day meeting at Caltech on 27 
July 1996.  This meeting, hosted by the Caltech Relativity Group, 
brought together, in person or by teleconference from Cardiff and 
Potsdam, many of the experts in the fields of post-Newtonian binary 
evolution calculations and numerical relativity for a discussion of 
these problems and possible approaches to their solution.

The meeting began with an overview by Richard Matzner,
principal-investigator of the Binary Black Hole Grand Challenge Team,
on the project status, followed by a presentation by Takashi Nakamura
on the on-going efforts in Japan to approach the same problem.
Discussion then turned, with presentations by Nakamura, Sasaki and
Wiseman, and by Seidel and Matzner, to the second question described
above: what is the minimum separation for which existing
post-Newtonian methods can give reliable results for a symmetric black
hole binary, and what is the maximum separation at which the numerical
calculations can begin if they are to carry the evolution reliably
through coalescence to the final state of a single, perturbed black
hole?

Several proposals were discussed for bridging the gap between the 
ending point of the reliable perturbative techniques used for the 
adiabatic inspiral and the fully numerical techniques being pursued 
for the coalescence.  Two of these proposals convey the range of 
options discussed.  Steve Detweiler described very promising work, 
just nearing completion, on a post-Minkowskii approximation scheme for 
iteratively constructing {\em spacetimes\/} (not spacetime slices) 
that satisfy the full field equations to fixed order in $G$.  On the 
other hand, Kip Thorne suggested that an adiabatic approximation to 
the field equations be sought that would allow the numerical solution 
to be carried out from larger separations.  One element of this 
approximation, which would deal with the ``dynamics'' associated with 
the motion of the black holes through the coordinate grid used in the 
numerical calculations (but not the dynamics associated with the 
physical propagation of radiation), is the use of a coordinate system 
that co-rotates with the binary.  Such a coordinate system introduces 
a light-cylinder, where the character of the coordinates change (some 
of the coordinates becoming light-like as one crosses the cylinder), 
and there was considerable discussion over the difficulties of 
handling this transition region, posing boundary conditions, and 
identifying the other components of the adiabatic approximation.

The discussion then turned briefly to the problem of identifying the 
numerical initial data for the coalescence calculations with a binary 
evolved from large separations.  Here, again, discussion covered the 
full range of possibilities.  Larry Kidder discussed a method under 
investigation with Sam Finn where the multipolar decomposition of the 
spatial metric and extrinsic curvature on a near-zone two-sphere 
surrounding the binary in the numerical initial data slice is compared 
to an identical decomposition of a similar slice through, {\em e.g.,} 
a post-Newtonian spacetime.  In the restricted context of binary black 
hole initial data and a point-mass binary post-Newtonian spacetime, 
intuition suggests that agreement of the moments with 
$\ell<\ell_{\max}$ suggests that the evolution of the numerical 
initial data represents an approximate continuation of the binary 
system evolved by post-Newtonian (or other) means from large 
separation, and that this approximation should become better as 
$\ell_{\max}$ increases.  The principle concern, voiced by Kip Thorne, 
is the identification of a prescription that identifies unambiguously 
equivalent two-spheres and multipole moments in the numerical initial 
data slice and the post-Newtonian spacetime.  On the other hand, Lee 
Lindblom suggested that if the evolution scheme used for the early, 
adiabatic inspiral could be made sufficiently accurate ({\em i.e.,} 
satisfy the constraints with sufficiently small residuals) at small 
separation, that a slice through the resulting spacetime could be used 
directly for as initial data for the fully numerical evolution, thus 
eliminating the ``seam'' that Kidder and Finn were attempting to sew.

Finally, Richard Price described on-going work with Andrew Abrahams,
Jorge Pullin and other collaborators on ``naive'' application of
perturbation theory to black hole coalescence.  Following-up on
earlier work by Abrahams and Cook, Abrahams, Price, Pullin and
collaborators use either the Zerilli equation for Schwarzschild
perturbations or the Teukolsky equation for perturbations of Kerr to
evolve the highly perturbed single black holes that exist immediately
following the formation of a single event horizon in a binary black
hole coalescence.  Doing so, they have found a remarkable and
unexpected agreement with the radiated energy of the fully numerical
simulation.  This work suggests that, for at least some purposes, the
validity of black hole linear perturbation theory may extend far into
the regime traditionally considered a large perturbation.

\eject
\section*{\centerline {Quantum Gravity in the Southern Cone}}
\addtocontents{toc}{\protect\medskip}
\addcontentsline{toc}{subsubsection}{\it Quantum Gravity in the Southern
Cone by Rodolfo Gambini}
\begin{center}
\medskip
Rodolfo Gambini,
Universidad de la Republica, Montevideo, Uruguay\\
\htmladdnormallink{rgambini@fisica.edu.uy}{mailto:rgambini@fisica.edu.uy}
\end{center}

The idea of this meeting was to bring together international researchers 
in quantum gravity with researchers from the area of the Southern Cone
of South America. The workshop was attended by 85 participants and took
place in Punta del Este, Uruguay on April 10-12 1996.

The plenary lectures included 
Esteban Calzetta speaking about Stochastic behavior in field theories
and semiclassical gravity, 
Jim Hartle on quantum cosmology and quantum mechanics, 
Marc Henneaux on cohomological methods in field theory, 
Gary Horowitz on black hole entropy in string theory, 
Carlos Kozameh on Fuzzy spacetimes, 
Karel Kucha\v{r} on quantum collapse, 
Juan Pablo Paz on decoherence, 
Jorge Pullin on knot theory and the dynamics of quantum gravity,
Carlo Rovelli on black hole radiation and entropy in loop quantum gravity
and
Lee Smolin on quantum spin networks and quantum gravity.

There were afternoon sessions including talks by

Max Ba\~nados, 
Mario Castagnino,
Alfredo Dominguez,
Hugo Fort,
Fabian Gaioli,
Jose Maluf,
Hugo Morales T\'ecotl,
Viktor Mostepanenko,
Javier Muniain,
Mike Ryan,
Victor Tapia,
Ranjeet Tate,
Thomas Thiemann,
Luis Urrutia,

and posters by 

Daniel Armand-Ugon,
Diego Dalvit,
Cayetano Di Bartolo,
Rafael Ferraro,
Fabian Gaioli,
Edgardo Garcia Alvarez,
Fernando Lombardo,
Daniel Sforza.

The conference was generally well received by the participants and
attracted a lot of coverage by the Uruguayan media. A second edition
of the conference will be organized in Bariloche, Argentina, in
January 1998.

\eject
\section*{\centerline{Report on the Spring APS Meeting}}
\addtocontents{toc}{\protect\medskip}
\addcontentsline{toc}{subsubsection}{\it 
Report on the Spring APS Meeting by Fred Raab and Beverly Berger}
\begin{center}
\medskip
Beverly Berger, Oakland University\\
\htmladdnormallink{berger@vela.oakland.edu}
{mailto:berger@vela.oakland.edu}
\end{center}

The Gravitation Topical Group (GTG) made its official debut at the
APS--AAPT Meeting in Indianapolis, 2--5 May 1996. Traditionally, this
meeting has enjoyed significant participation by the Divisions of
Astrophysics (DAP), Particles and Field (DPF), and Nuclear
Physics. Decades ago, there were also several sessions on
gravitational physics. This participation had declined over the years
but, with the formation of the GTG, has now experienced a strong
revival. The GTG sponsored a well-attended invited session with talks
by Cliff Will (``Gravitational Waves and the Death-Dance of Compact
Stellar Binaries''), Fred Raab (``Progress Toward a Laser
Interferometer Gravitational Wave Observatory''), Ho Jung Paik
(``Spheres---Omni-directional Multi-mode Gravitational Wave Antennas
for Next Generation''), and Matt Choptuik (``Critical Phenomena in
Gravitational Collapse''). There were also two joint invited
sessions. The first with the Topical Group on Fundamental Constants
and Precision Measurements (FCTG) featured talks by Francis Everitt
(``From Cavendish to the Space Age: Some Thoughts on the History of
Precision Measurements''), Jim Faller (``Precision Measurements with
Gravity''), Riley Newman (``New Measurements of $G$''), and Paul
Worden (``Testing the Equivalence Principle in Space''). This session
was so successful that FCTG and GTG will co-sponsor another session at
the 1997 Spring Meeting. The other joint session with DAP focused on
neutron stars with talks by Peter Meszaros (``Neutron Star Models and
Gamma Ray Bursts''), Dong Lai (``Learning about Neutron Star from
Coalescing Compact Binaries and Radio Pulsar Binaries''), John
Friedman (``General Relativistic Instabilities of Neutron Stars''),
and Charles Meegan (``Observations of Gamma Ray Bursts'').  There were
also a number of contributed papers that were divided among three
sessions: Numerical Relativity, Black Holes, and Cosmology (chaired by
Matt Choptuik), Gravity Experiments and Theory (chaired by Fred Raab),
and Gravitation Theories (chaired by David Garfinkle). The GTG also
held its first business meeting at the conference. In addition to
these official GTG activities, there was other evidence of the
vitality of gravitational physics. A special plenary session of the
APS featured Kip Thorne's Lilienfeld Prize Lecture (``Black Holes,
Gravitational Waves, and Quantum Non-Demolition'') while a joint
Division of Particles and Fields-DAP session on Particle Astrophysics
included an invited talk by Barry Barrish (``The Detection of
Gravitational Waves''). Finally, this interest in LIGO provided a
backdrop for the meeting of the LIGO Research Community which will
also participate in the 1997 Meeting.

Details and abstracts can be found at
\htmladdnormallink{\protect {\tt 
http://www.aps.org/BAPSMAY96/index.html}}
{http://www.aps.org/BAPSMAY96/index.html}. 
For those who could not attend the
meeting, the minutes are given below.

Minutes for Business Meeting of Topical Group on Gravitation

Executive Committee members present:  Berger, Thorne, Bardeen, Parker, Raab,
Shoemaker, Finn. Absent:  Ashtekar, Isenberg, Wald

The meeting was called to order at approximately 5 pm by Beverly Berger.

Beverly Berger introduced the officers of the topical group and gave
a brief description of the membership statistics. The topical group
currently has approximately 300 members, significantly above the
minimum level of 200 members required to form and maintain such a
group.  It was noted that there are approximately 150 people who
signed the petition requesting formation of the topical group but are
not yet members. Efforts will be made to bring these people into the
group's membership. The membership is comprised of approximately equal
numbers of theoretical and experimental investigators. 

The status of committees was reported. The nominating committee, chaired by
David Shoemaker, will begin to prepare for election of a vice-chair and two
executive-committee members this autumn. The Fellowship Committee, chaired by
Abhay Ashtekar, is currently considering nominations for APS Fellowship. 

A report was given on the state of the group's finances (as provided
by Jim Isenberg). Income, principally from membership fees, totalled
\$1417. Expenses, associated with printing and mailing the newsletter,
``Matters of Gravity'', totalled \$943. This leaves a balance of \$473
in the Treasury. The issue of potential cost savings by using
electronic distribution of ``Matters of Gravity" was raised. Members
in attendance voiced agreement with Isenberg's suggestion that future
distributions of the newsletter would be done electronically as far as
possible, provided that members could still opt for a paper copy if
electronic access presented problems. Members would be contacted by
e-mail concerning whether they want electronic or paper copies of the
newsletter in future.  Attendees at the meeting were asked for
suggestions of how these funds might be used to good
effect. Suggestions were made that something to encourage student
participation, either through support for attending meetings or an
award, might be a good use for funds.

Beverly Berger advised the audience that organizing future meetings
would be an important issue in the near future. The precise details
were not yet clear, because the first Meetings Committee of APS to
involve the Topical Group on Gravitation would only meet later in the
week. Anticipating that our small topical group would get only a few
meetings slots at the next April meeting, the general sentiment
supported splitting those slots with other groups that had shared
interests. This had worked well at this meeting and was thought to
provide better exposure with limited speaking slots. The issue of joint
sponsorship of gravitational physics meetings that already occur on a
periodic basis was raised, but further work was needed to identify what
the APS rules are in this area.

Beverly Berger Adjourned the meeting at approximately 5:40 pm.

Minutes submitted by Fred Raab
\htmladdnormallink{fjr@ligo.caltech.edu}
{mailto:fjr@ligo.caltech.edu}, 
(with slight revisions by Beverly Berger).

\end{document}